# DREAM_OLSR PROTOCOL (Distance Routing Effective Algorithm for Mobility - Optimized Link State Routing)

Gurpreet Singh Saini[#1], Ashish Kots[#2], Manoj Kumar[#3]

[#]*Amity School of Engineering and Technology, Amity University*
*Sec-125, NOIDA, Uttar Pradesh, India*

***Abstract* - This paper lays down a proposal of protocol named DREAM_OLSR. The protocol has been developed so as to effect current OLSR (RFC 3626) [4] protocol. The protocol establishes an optimized solution hence the name has been manipulated from Open Link State Routing to DREAM Optimized Link State Routing. DREAM specifies Distance Routing Effective Algorithm for Mobility wherein it implements the Distance routing effective algorithm for the optimized solution. This optimization includes higher efficiency and fewer overheads for the MANET.**

***Keywords*— MANET, OLSR, Routing, Ad-hoc, MPR**

## I. Introduction

In today's scenario where mobility is pre-requisite a need of mobile channels has been raised to keep connections alive. And therefore raises a need for protocol which can give us mobility. Here MANET acts as a channel. MANET's are self-configuring networks where every node is free to move. But, the main problem lies with selection of optimum protocol to lure best services from the channel. To this problem a proposal of protocol DREAM-OLSR which is an extension to the current OLSR routing protocol (RFC 3626) is presented here.

MANET's are self-configuring in nature and that nature has been supported by three basic protocol standards:

- AODV (Ad-Hoc On Demand Distance Vector) RFC 3561[5]
- DSR (Dynamic Source Routing) RFC 4724[6]
- OLSR (Open Link State Routing) RFC 3626

The DREAM_OLSR inhibiting some specific properties optimization OLSR is presented here. Hence, minimising costs and achieving higher throughput while having following characteristics.

- Proactive & Table-driven.
- Based on the link state algorithm [4].
- When a node or link changes its state, its information is updated [2].
- Nodes in the network share their information such that nodes present in the network can construct a map of the whole network using the Multi Point Relay (MPR) Selection[2][3].
- It uses the MPR nodes to share information in the network and by this way optimized the flooding [2] [3].
- Performs hop by hop routing [1] and maintains routes for the various nodes based on the dynamic table entries.

## II. Design

There are various components in DREAM_OLSR protocol. The design [11] is based on link state routing principal**.** The main focus of this paper is to present protocol in such a way that it is Optimized, more secure and reliable.

## III. Data Structures Used By DREAM_OLSR

DREAM_OLSR requires data structures basically for maintaining itself and for loading & fetching information and uses following:

*A.* *Link Set:* It is based on interface-to-interface links concept and used to calculate the present state of links.

*B.* *Neighbor Set*: All neighboring registered nodes are monitored here. Information about the nodes is updated and configured dynamically.

*C.* *Multiple Interface Association Information Base:* This holds info of nodes using more than one interface for communication with other nodes.

*D.* *2-hop Neighbor Set:* It contains the information about nodes that have path through a one–hop neighbor.

*E.* *MPR Set:* It contains information about all the MPR's selected by their local nodes.

*F.* *MPR Selector Set:* This contains link information of all the neighboring nodes that have a common mutual agreement.

*G.* *Topology Information Base:* This is used to maintain information of dynamically changing nodes & information of all link-state information received from the other nodes in the MPR set.

*H.* *Duplicate set:* It is used to store information about the messages that are being recently processed.

*I.* *Timeouts:* Timeouts are used to indicate the validity of the registered information. Entries are removed if they miss timeouts.

## IV. PACKET FORMAT

DREAM_OLSR uses packets for information exchange consisting; a header and a functional layout as shown in below figure:

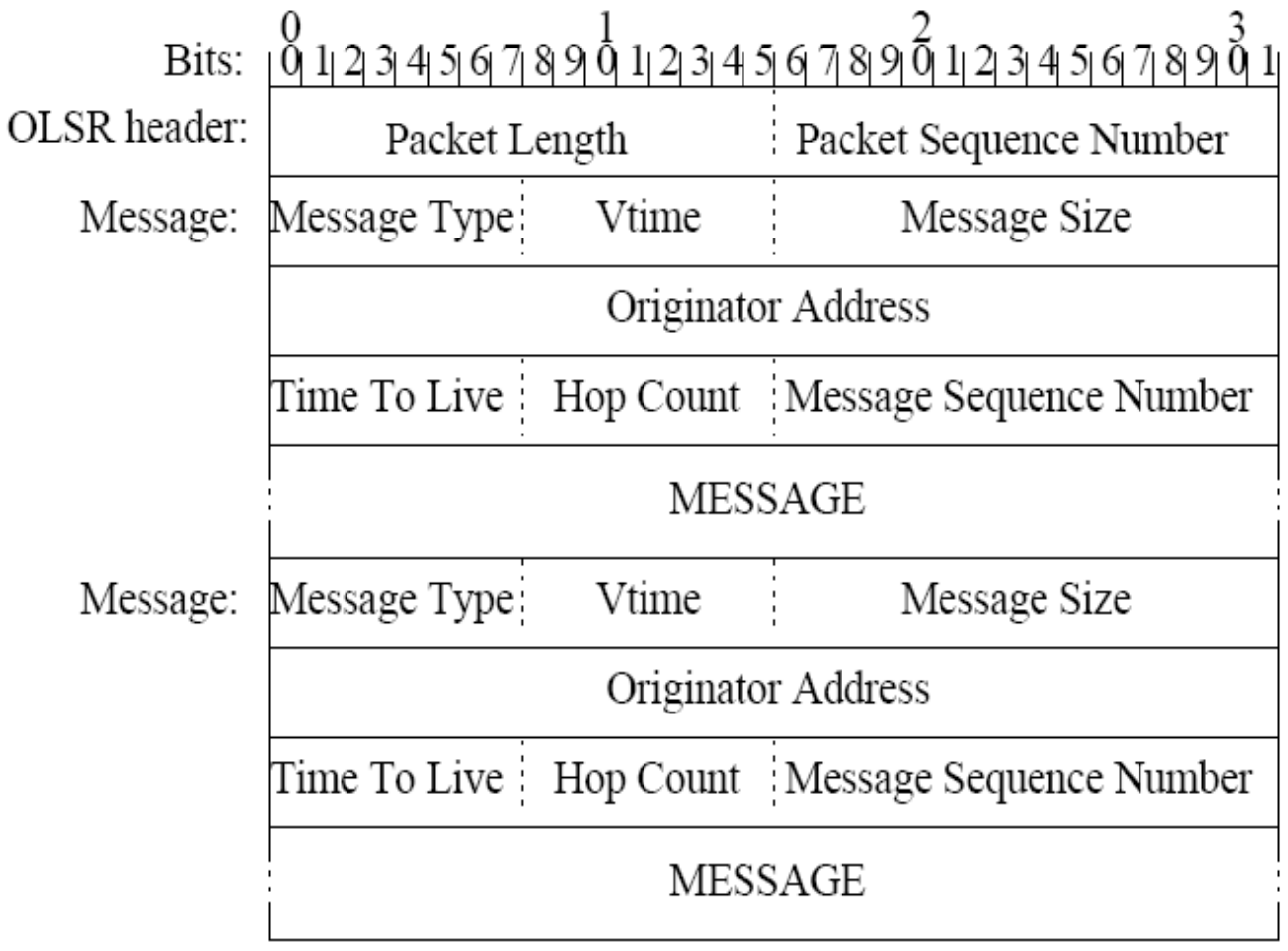

Fig 1: Packet format of DREAM_OLSR

A. *Packet Length:* The size of whole packet, including everything header, message etc.

B. *Packet Sequence Number*: Whenever a host transmits a message its sequence number is incremented by one.

C. *Message type*: DREAM_OLSR uses an integer value for identifying the message type and has reserved message types of 0-127. 128-255 space is considered “private” and plays an important role in the custom extension of protocol.

D. *V-Time:* It contains information about validation of information contained by a message in Mantissa exponent format.

E. *Message Size*: No. of bytes contained by message, including message header.

F. *Originator Address*: Address of the node originating the message.

G. *Time to Live*: It indicates max number of hops up-to which a message can be transferred, used to control the flooding.

H. *Hop Count:* It contain number of hops a message has passed.

I. *Message Sequence Number:* Every new packet in DREAM_OLSR has a unique sequence no.

## V. MULTIPOINT RELAYS

In Broadcasting, to reduce the number of duplicate retransmissions of packets, DREAM_OLSR uses concept of MPR [4] through which flooding of packets is controlled. MPR’s are selected in a way that all 2 hop neighbours can be reached through a MPR. All the black nodes in the below figure represents the concept of MPR. It’s clear that all two hop nodes in the network can be reached through a MPR.

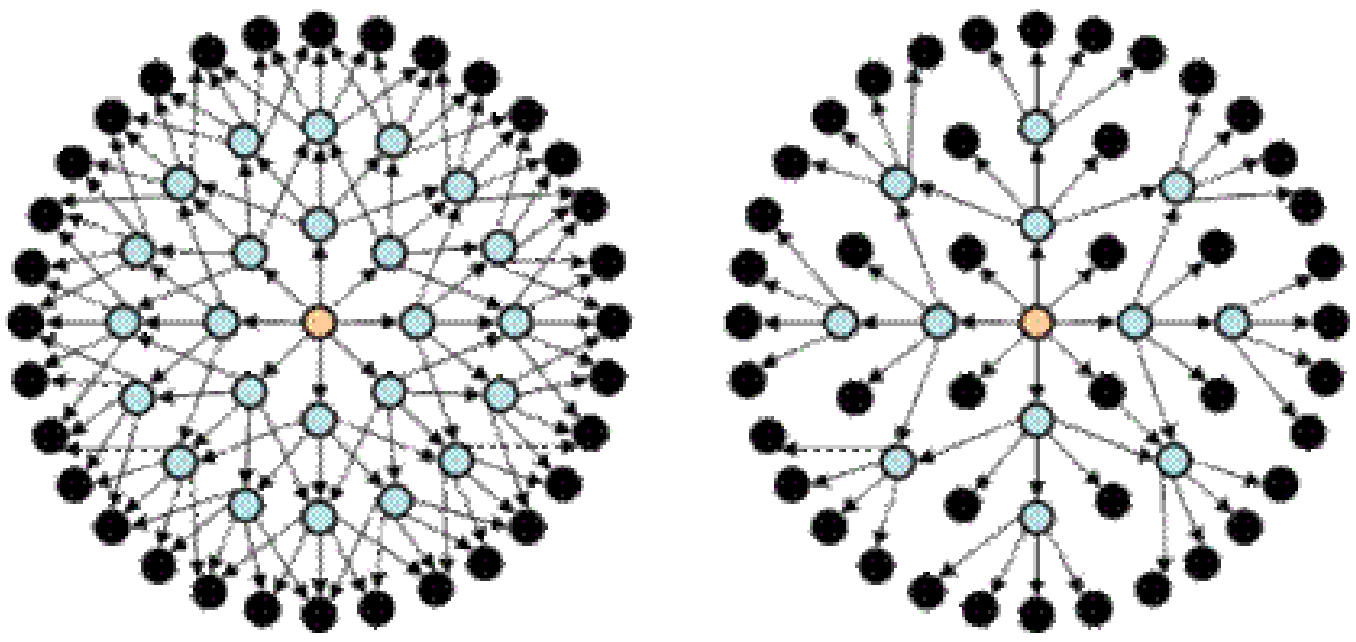

Fig 2: a) A wireless multi hop network for flooding a packet.

b) A wireless multi hop network using MPRs for flooding.

## VI. TRAFFIC FORWARDING IN DREAM_OLSR

MANET uses the concept of relaying of message for flooding. To flood packets in network DREAM_OLSR first defines a forwarding algorithm and uses the updated MPR information. However every node in the network is free to make its own rules for custom forwarding of messages.

The algorithm can be presented as follows:

- If the message arriving is not symmetric, it is discarded and the link set is checked for the link status.
- If the header of TTL message is 0, then also it is discarded.
- The message already forwarded is discarded after verification from duplicate set.
- MPR selector set is used to check whether the last hop sender belongs to it or not, if not then message is discarded.
- Whenever a message has to be forwarded its hop count is increased by one and TTL is decreased by one before broadcasting.

## VII. LINK SET OPTIMIZATION

The nodes chosen as MPR by their neighbors have to declare information about their link-state and needs to declare info about the MPR selectors in the link-state messages. By flooding the information in the network nodes calculate shortest path routes to all the hosts. By Default only the node which has at-least one neighbor selecting it as MPR; only floods link-state messages wherein messages they include MPR selectors.

## VIII. NEIGHBOR DISCOVERY

This process is dependent upon HELLO messages which are emitted at regular intervals. The discovery is done as follows:

- Node A sends an empty HELLO message. Upon receive of this message Node B registers A as an asymmetric neighbor as message not includes its address.
- Then B sends HELLO message declaring A as an asymmetric neighbor.
- When A receives this message, it sets B as symmetric neighbor as message contains its address.
- Finally A includes address of B in HELLO and sends a message back which notifies B to add an as symmetric neighbor.

Link, neighbor, 2-hop neighbor & MPR selector sensing is done by the use of HELLO messages. In HELLO message, MPR nodes transfer all known links and neighbors. The Links and neighbors are registered and grouped to optimize byte usage. For every interface HELLO messages are generated so as to beneficiate link sensing, which makes use of possible non-main-addresses.

## IX. NEIGHBOR AND MPR SELECTOR SET

A node maintains a repository of the one-hop neighbors using main addresses of nodes. The neighbor entries directly relate to link entries. The neighbor table is queried whenever a link entry is created. Only one neighbor entry exists per neighbor. The entries of neighbors are manipulated as per change in link-set. Node also contains information about all other nodes that have a path through symmetric neighbors. Neighbor set contains information about the two hop nodes and this is used to calculate multi point relays. Whenever a node received a HELLO message from its symmetric neighbor, information about all symmetric neighbors is updated or added in the neighbor set of two hop neighbor. Every node entry in the two hop neighbor set is based on main addresses and every node makes a query to the MID set for main address.

Flooding scheme in the multi-point relays is based on the registered nodes & the neighbors that have selected them as MPR. MPR_NEIGH is used to match the address & the neighbor type. If a new entry is found there is an update in the MPR selector set using the senders HELLO message.

## X. LINK STATE DECLARATION

The concept of flooding is used by Link state routing protocols to share information about their local links. It also uses the concept of host based flat routing, for this Topology Control messages are used.

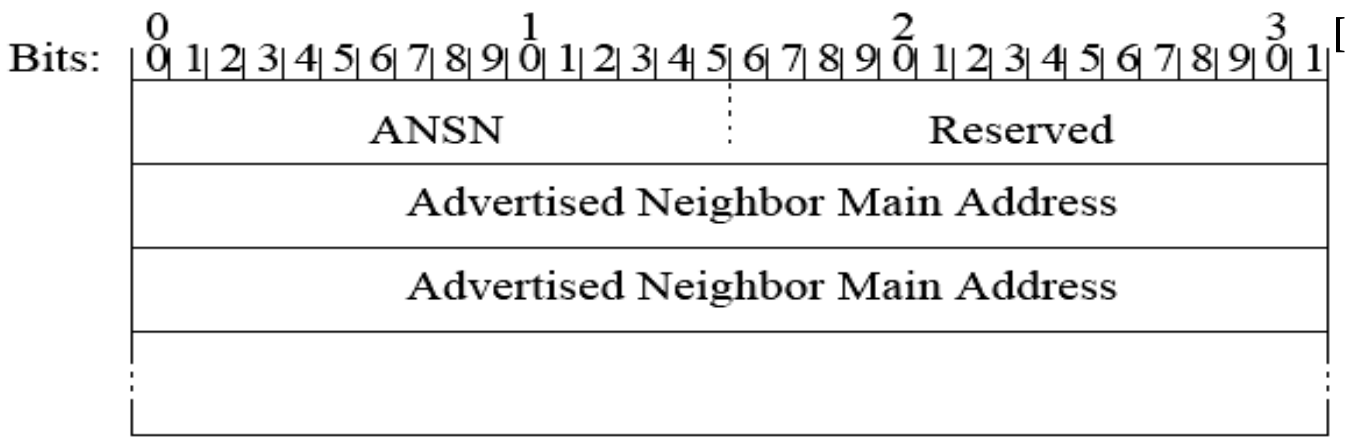

Fig 3: The OLSR Topology Control message format

## XI. SIZE OPTIMIZATION

Size optimization is also taken care of as the size of TC message is optimized because a node only declares its MPR selectors in respective TC message. This reduction is directly proportional to the density of the network topology under consideration. In this topology, size of the original TC message for the central node would become half.

## XII. SENDER OPTIMIZATION

The nodes not falling under MPR set does not broadcast any messages or status updates. Also, it includes the nodes which lost their status of MPR nodes due to dynamic configurations. However, these nodes do generate TC messages which do not contain any information and are just used for the configuration purposes. In real time scenario only the nodes other than above mentioned generate specific TC messages so as to create MPR selector sets reducing overheads hence optimizing.

## XIII. ROUTE CALCULATION

Many heuristics have been proposed for route calculation [13] in RFC3626 and DREAM_OLSR uses same for the discovery of most efficient shortest-path algorithms. These can be outlined as:

*A.* All one hop symmetric neighbors are registered in tables at MPR set nodes with a hop-count of 1.

*B.* For each symmetric one-hop neighbor, neighbors that have not been added in the routing table & have max two hop neighbor distances and have maintained a symmetric link to that neighbor. The entries are made with two hop-count and next-hop as the current neighbor.

*C.* Then, for every node N parsed into the routing table with hop-count of 2 enter all entries from the TC set which includes:

*D.* Information about the originator of the packet in the Topology control entry =N

*E.* All the entries that come later will be added with n+1 hop count.

*F.* Increase the value of N with 1 and repeat the step 3 until routing table contains no entries with a hop count of N+1.

*G.* MID set is queried for every entry E for address aliases. If such address aliases exist a new entry is updated in the routing table with hop count set to Es.

## XIV. DREAM_OLSR INFORMATION DATA STRUCTURES

DREAM_OLSR has three main modules over which it works dependent over the functions they perform namely neighbor sensing, MPR selection and Link State Flooding. It has also been observed that most of the traffic is generated dependent over the set of repositories maintained in the routing tables and are updated dynamically based upon the received control messages transferred.

Figure 4 depicts how information repositories in DREAM_OLSR are maintained and how it relates to other features. Updates in the link set is done after receiving Hello messages from either existing or new node which then initiates updates in the neighbor set, which in-turn again initiates recalculation of the MPR selector set. TC messages initiates updating in the topology set. If the received messages are not recorded or registered, they are updated in Duplicate set. To generate a HELLO message it requires the link set, neighbor set and MPR set. The generated TC messages are derived from MPR selector set and the Duplicate set.

After fetching information from neighbor, 2-Hop neighbor and traffic control, route calculation is done. The GUI and other applications can use these structures for their calculations by fetching these parameters into their local structures and databases.

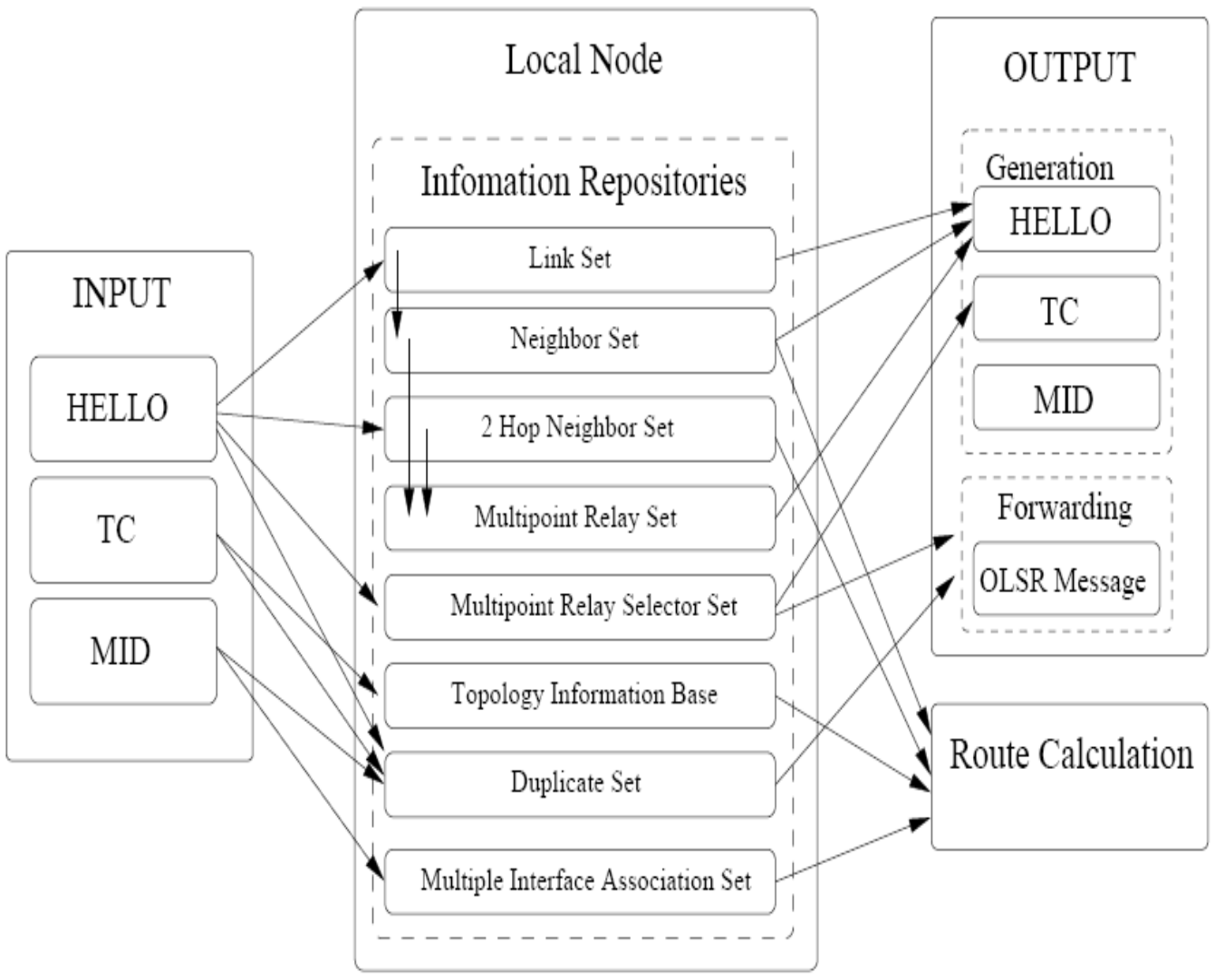

Fig 4: Information Repositories of OLSR relation

## XV. RESULTS

### *A. Parameters*

The tool used for the simulation results is ns version 2.27 over the Red Hat Linux 9.0 and under the Random Waypoint Model where nodes move in 1000m by 1000m area. The data speed is 2mbps and the TC and HELLO message interval is set to be 0.5 seconds and data size is 64 bit.

### *B. Simulation Evaluation Metrics*

I. Average throughput
II. Packet Delivery Ratio
III. Packet Delay

### *C. Outcomes*

The results shown below show the effect of mobility, MPR's selected per network and no. of retransmissions of the TC messages in the network. The fig 5 explains how the increase in number of MPR's in the network increase the throughput of the network. More the MPR's, better the throughput. This enhances the capability of the DREAM_OLSR in comparison to previous versions.

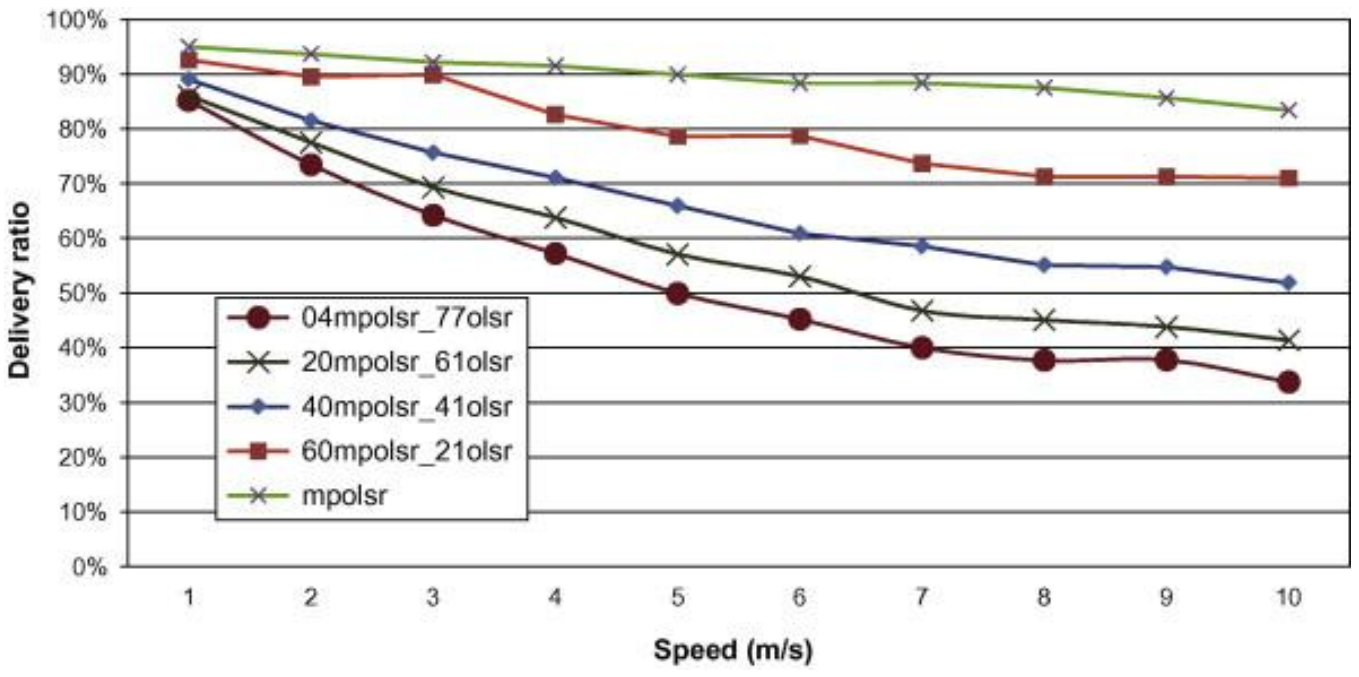

Fig 5: Throughput dependent on MPR

The throughput however does increases with the number of the MPR's selected in the network but it does increase the overheads in turn as every MPR node maintains a complete table of the network. Also, this leads to delay of data in network as every MPR node recalculates the effective shortest path using algorithm. The following figure 6 marks the increase in delay of the data as the number of MPR's increase.

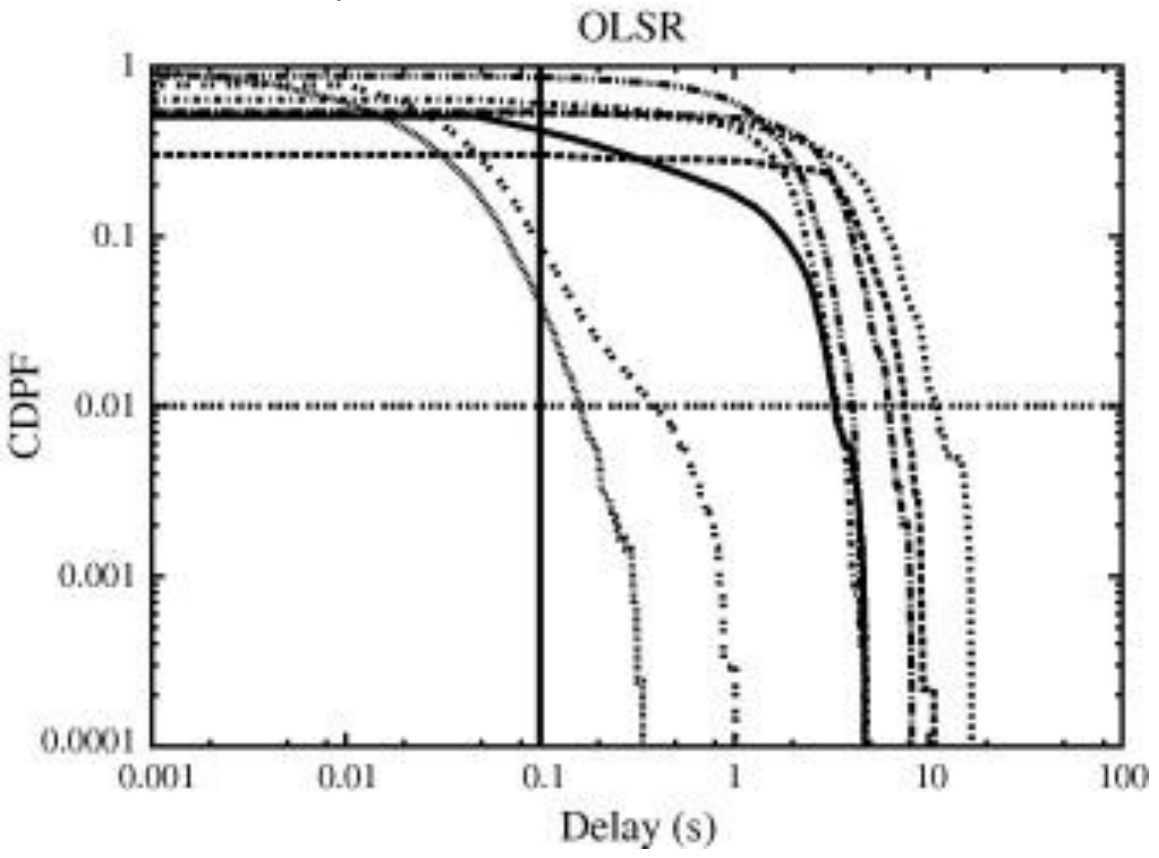

Fig 6: Delay in network

To produce efficient results the number of TC retransmissions need to be varied and according to number of mobile nodes in the network. DREAM_OLSR produces the following result. Figure 7 marks how the TC retransmissions are vary in the compared three flavours and also mark how DREAM_OLSR performs efficiently by decreasing the number of TC transmissions.

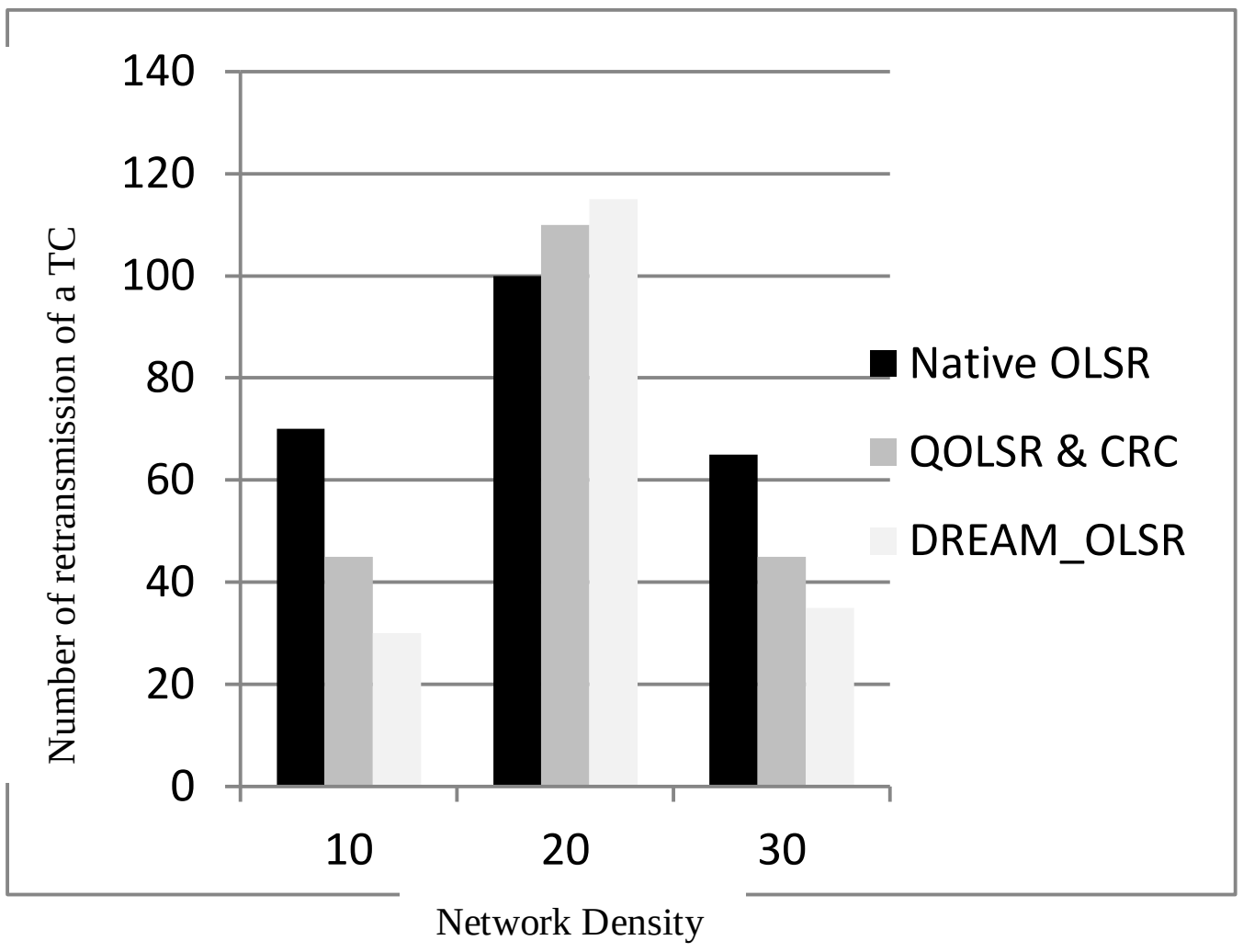

Fig 7: Network Density Comparison

## XVI. Future Scope

This paper covers the modification in the procedure of MPR selection along with the effect of the number of the MPR nodes in the MPR selector set. The refinement is thus achievable but depending upon the network, there raises a need of special mechanism to automate network for artificial learning and intelligence mechanism wherein it can decide the optimum sets. There is a complete need for refinement as the DREAM_OLSR seems to exhaust when the number of nodes in network are too high (> 102400). The security feature also needs to be looked upon depending upon the data to be transmitted.

## XVII. Conclusion

MANETS are need of today; the optimization defines a step towards working in a direction to make it more feasible for the deployment in practical scenarios. Here DREAM_OLSR provided better results as compared to present solutions with some boundaries.

## References

[1] S. Kulasekaran and M. Ramkumar, "APALLS: A Secure MANET Routing Protocol", January 2003.

[2] Xin Wang, "Mobile Ad-Hoc Networks: Application", January 2011.

[3] Xin Wang, "Mobile Ad-Hoc Networks: Protocols Design", January 2011.

[4] T. Clausen and P. Jacquet, "Optimized Link State Routing Protocol," RFC 3626, October 2003.

[5] C. Perkins, E. M. Royer and S. R. Das, "Ad Hoc On Demand Distance Vector (AODV) routing," RFC 3561, July 2003.

[6] D. Johnson and Al, "The Dynamic Source Routing Protocol for Mobile Ad hoc Networks (DSR)," In IETF Internet Draft, draft-ietf-manet-dsr-09.txt, April 2003.

[7] http://www.ietf.org/rfc/rfc3626.txt

[8] http://www.ietf.org/rfc/rfc3561.txt

[9] http://www.ietf.org/rfc/rfc4728.txt

[10] http://www.cse.yorku.ca/course_archive/2011-12/F/3213/ Notes/chapter _19_bellman.Pdf

[11] http://www.omnisecu.com/cisco-certified-network-associate-ccna/ introduction-to-link-state-routing-protocols.htm

[12] D. Johnson and G. Hancke, "Comparison of Two Routing Metrics in OLSR on a Grid based Mesh Network", Ad Hoc Networks, Vol. 7, No. 2, pp. 374-387, March 2009.

[13] Luo Junhai, Ye Danxia, Xue Liu, And Fan Mingyu, A Survey Of Multicast Routing Protocols For Mobile Ad-Hoc Networks, IEEE Communications Surveys & Tutorials, Vol. 11, No. 1, First Quarter 2009, Pp:78-91.

[14] Nadir Shah, Depei Qian, Khalid Iqbal, Performance Evaluation Of Multiple Routing Protocols Using Multiple Mobility Models For Mobile Ad Hoc Networks, Proceedings Of the 12th IEEE International Multi topic Conference, December 23-24,2008, Pp:243-248.